\documentclass{elsarticle}
\usepackage{graphicx}
\usepackage[intlimits]{amsmath}
\usepackage{amssymb,amsfonts}
\usepackage{bbm}
\usepackage{booktabs}
\usepackage{mathrsfs}
\usepackage{textcomp}
\usepackage[T1]{fontenc}
\usepackage{lmodern}

\renewcommand{\vec}[1]{\boldsymbol{#1}}

\newcommand{\ud}{\text{d}}
\newcommand{\ex}{\text{e}}
\newcommand{\im}{\text{i}}
\newcommand{\id}{\mathbbm{1}}

\begin{document}

\begin{frontmatter}

\title{A relativistic description of the A$(\pi^+,K^+)_{\Lambda}$A reaction}

\author[Giessen]{S. Bender}
\author[Giessen,Saha]{R. Shyam}
\author[Giessen]{H. Lenske}

\address[Giessen]{Institut f\"ur Theoretische Physik, Universit\"at Giessen,
D-35392 Giessen, Germany}
\address[Saha]{Saha Institute of Nuclear Physics, Kolkata 700064,  India.}

\begin{abstract}
We investigate the A$(\pi^+,K^+){_\Lambda}$A reaction within a covariant
model. We consider those  amplitudes which are described by creation,
propagation and decay into relevant channel of $N^*$(1650), $N^*$(1710),
and $N^*$(1720) intermediate baryonic resonance states in the initial
collision of the incoming pion with one of the target nucleons. The bound
state nucleon and hyperon wave functions are obtained by solving the Dirac
equation with appropriate scalar and vector potentials. Expressions for the
reaction amplitudes are derived taking continuum particle wave function in
the plane wave approximation. Numerical calculations are presented for
reactions on $^{12}$C, $^{40}$Ca, $^{51}$V and $^{89}$Y target nuclei. The
predictions of our model are in reasonable agreement with the available
experimental data.
\end{abstract}


\end{frontmatter}

\section{Introduction}
Hypernuclei, where one or two nucleons ($N$) are replaced by hyperons
($Y$) in the bound orbits, provide a unique opportunity for studying
a new form of the hadronic system which has strangeness degrees of
freedom~\cite{pov87,chr89,ban90,has06}. Lambda ($\Lambda$) hypernuclei
are the most familiar and extensively investigated hypernuclear systems.
Since the $\Lambda$ hyperon does not suffer from Pauli blocking by other
nucleons, it can penetrate deep inside the nucleus and form deeply bound
hypernuclear states. Thus, hypernuclei can provide information about the
nuclear states which are not accessible in ordinary nuclei. Such systems
are perhaps the only tool currently available to get information about the
$\Lambda$-$N$ interaction as $\Lambda$-nucleon scattering experiments
are very difficult to perform due to short life time of the $\Lambda$
particle. During the past years, data on hypernuclear spectroscopy have been
used extensively to extract information about the hyperon-nucleon
interaction within a variety of theoretical approaches (see, e.g.,
Refs.~\cite{hiy00a,hiy00b,hiy00c,hiy00d,hiy00e,Tsushima:1997cu,kei00a,kei00b}).

$\Lambda$ hypernuclei have been studied extensively by the stopped as well
as the in-flight $(K^-,\pi^-)$ reaction (see, e.g., the
reviews~\cite{chr89,ban90,has06,may81a,may81b,may81c})
and also by the $(\pi^+,K^+)$ reaction~\cite{pil91a,pil91b,hot01a,hot01b}.
The kinematical properties of
the $(K^-,\pi^-)$ reaction allow only a small momentum transfer to the
nucleus (at forward angles), thus there is a large probability of populating
$\Lambda$-substitutional states ($\Lambda$ assumes the same orbital angular
momentum as that of the neutron being replaced by it). On the other hand,
in the $(\pi^+,K^+)$ reaction the momentum transfer is larger than the
nuclear Fermi momentum. Therefore, this reaction can populate states
with the configuration of an outer neutron hole and a $\Lambda$ hyperon
in a series of orbits covering all the bound states having
high spin natural parity configurations. The richness of the
spectroscopic information on $\Lambda$ bound states in the
$(\pi^+,K^+)$ reaction was demonstrated in the
experiments performed at the Brookhaven National Lab and National
Laboratory for High Energy Physics (KEK) (see, e.g.,
Ref.~\cite{has06} for a comprehensive review).
Furthermore, although the reaction cross section of the strangeness
production, via the $(\pi^+,K^+)$ process, is smaller than that of the
strangeness exchange reaction $(K^-,\pi^-)$, the higher luminosity of
pion beams makes experiments more feasible.

In the experimental studies reported in Refs.~\cite{hot01a,hot01b}, this reaction has been used
to carry out the  spectroscopic investigations of hypernuclei ranging
from light mass $^{12}_{\Lambda}$C, to medium mass $^{51}_{\Lambda}$V and
$^{89}_{\Lambda}$Y with the best resolution (${\sim}1.6$--$1.7\,$MeV) achieved in the
spectrometer at KEK. This experiment has succeeded in clearly observing
a characteristic fine structure in heavy systems by precisely obtaining a
series of $\Lambda$ single-particle states in a wide range of excitation
energies.

Most of the  theoretical models used so far to describe the $(\pi^+,K^+)$
reaction employ a non-relativistic distorted wave impulse approximation (DWIA)
framework~\cite{dov80} (see also Ref.~\cite{ban90} for a comprehensive review of
these models). In these calculations, the $\Lambda$ bound states are generated
by solving the Schr\"odinger equation with Woods--Saxon or harmonic oscillator
potentials. However, for processes involving momentum transfers of
typically $300\,$MeV/c or more, a non-relativistic treatment of the corresponding
wave functions may not be adequate as in this region the lower component of the
Dirac spinor is no longer negligible in comparison to its upper component
(see, e.g., Ref.~\cite{shy95}).

\begin{figure}
\centering
	\includegraphics[width=.15\textwidth]{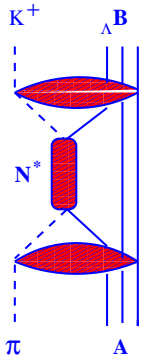}
\caption{
Graphical representation of our model to describe the $(\pi^+,K^+)$ reaction.
The elliptic shaded area represent the optical model interactions in the
incoming and outgoing channels.
}
\label{fig:Fig1}
\end{figure}
In this paper, we study the A$(\pi^+,K^+){_{\Lambda}}$A reaction within a
fully covariant model by retaining the field theoretical structure of the
interaction vertices and by treating the baryons as Dirac particles. In
this model, the kaon production proceeds via the collision of the projectile
pion with one of the target nucleons. This excites intermediate baryon
resonance states ($N^*$) which decay into a kaon and a $\Lambda$ hyperon.
The hyperon is captured in the respective nuclear
orbit while the kaon rescatters onto its mass shell (see Fig.~\ref{fig:Fig1}).
A similar picture has been used to describe the $A(p,K^+){_{\Lambda}}B$
and $A(\gamma,K^+){_{\Lambda}}B$ reactions in Refs.~\cite{shy04,shy06,shy08}.
In our model, the intermediate resonance states included are
$N^*$(1650)[$\frac{1}{2}^-$], $N^*$(1710)[$\frac{1}{2}^+$], and
$N^*$(1720)[$\frac{3}{2}^+$] which have dominant branching ratios for
the decay to the $ K^+ \Lambda$ channel~\cite{shy99,PDG2008}. Terms corresponding to
the interference among various resonance excitations are included in the
total reaction amplitude.

In section~\ref{chap:model}, we present the details of our formalism for calculating
the amplitudes corresponding to the diagrams shown in figure~\ref{fig:Fig1}.
In section~\ref{chap:results}, numerical results are presented for the $(\pi^+,K^+)$
reaction on $^{12}$C, $^{40}$Ca, $^{51}$V and $^{89}$Y targets using
continuum wave functions in the plane wave approximation. Summary, conclusions
and future outlook of our work are given in section~\ref{chap:summary}.

\section{Covariant model for the A$(\pi^+,K^+)_{\Lambda}$A reaction}
\label{chap:model}

\begin{figure}
\centering
\includegraphics{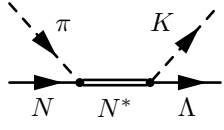}
\caption{Tree diagram for the elementary process of pion-induced strangeness
	production via resonance excitation and decay on a single nucleon.}
\label{fig:Fig2}
\end{figure}

The structure of our model for the $(\pi^+,K^+)$ reaction is described in
Fig.~\ref{fig:Fig1}. The $N^*$ corresponds to
the $N^*(1650)[\frac{1}{2}^-]$, $N^*(1710)[\frac{1}{2}^+]$,
and $N^*(1720)[\frac{3}{2}^+]$ baryon resonance intermediate states.
Terms corresponding to interference between various amplitudes are retained.
The elementary process involved in this reaction is shown in Fig.~\ref{fig:Fig2}.

It is clear that our model has only $s$-channel resonance contributions. In
principle, Born terms and the resonance contributions in $u$- and $t$-channels
should also be included in description of both the processes depicted by
Figs.~\ref{fig:Fig1} and~\ref{fig:Fig2}.
These graphs constitute the non-resonant background terms. Their
magnitudes depend on particular models used to calculate them and also the
parameters used in those models.
This can be seen in Ref.~\cite{feus98} where the effect of the background terms
is studied for $S_{11}$ phase shifts in the pion-nucleon interaction. Whereas
the contributions of the background terms are about $15$--$20\,\%$ of the resonance
terms within a coupled-channel $K$ matrix model, they are limited to less than
$10\,\%$ in the model of Ref.~\cite{dyt97} for the energies of our interest
($\approx 1\,$GeV/nucleon; the corresponding invariant mass is about $1.7\,$GeV) (see
Fig.~15 of~\cite{feus98}).  The effect of the background terms on the total
production cross sections of the $\pi^- p \to K^0 \Lambda$ reaction at our beam
energies can be indirectly inferred from the calculations reported in
Ref.~\cite{shk05}.  It is seen in Fig.~2 of this reference that the total
production cross sections at invariant mass around $1.7\,$GeV are dominated by the
contributions of $S_{11}$ and $P_{11}$ resonance terms. Therefore, the
magnitudes of the background terms are likely to be limited to about $10$--$20\,\%$
of those of the resonance terms.  Our results reported later on in this paper
may be uncertain by $10$--$20\,\%$ due to omission of the background contributions.

\subsection{Interaction Lagrangians}\label{sec:lagrange}

For the interaction terms of the spin-$1/2$ resonances, we have
vertices of \emph{pseudoscalar} (PS) or \emph{pseudovector} (PV) form.
The pseudovector coupling is consistent with the chiral symmetry
requirement of the fundamental theory of strong interactions
(quantum chromodynamics (QCD)).
In contrast to that, the pseudoscalar one does not have this property,
but it is easier to calculate.
The couplings are in both cases fixed in such a way that they are equal on-shell;
for off-shell cases, their difference is suppressed due to the denominator of the
resonance propagator.
It is, therefore, arguable which Lagrangian to use.
The best approach would be to introduce a mixing parameter, which was investigated
in Refs.~\cite{Gridnev1999,shy08}.

To avoid the introduction of additional parameters in our model due to a PS-PV
mixing for the interaction Lagrangians of Refs.~\cite{Gridnev1999,shy08},
we use the convention of either choosing the PS or the PV
couplings for these vertices as done in Refs.~\cite{feus98, Feuster1999}.
This is in line with the studies reported in Refs.~\cite{Sauermann1995,pen02}.

The pseudoscalar interaction Lagrangians for the spin-$1/2$
resonances are given by
\begin{subequations}\label{eq:lag.ps}
\begin{align}
	\mathcal{L}_{\pi N N_{1/2}^*}^{\text{PS}} & = - g_{\pi N N^*}
		\bar{\psi}_{N^*} \Gamma ( \vec{\tau} \cdot \vec{\phi}_{\pi} ) \psi_{N}
		+ \text{h.~c.}
	\label{eq:l.ps.piNR} \;, \\
	\mathcal{L}_{N_{1/2}^* K\Lambda}^{\text{PS}} & = - g_{N^* \Lambda K}
		\bar{\psi}_{N^*} \Gamma \phi_{K} \psi_{\Lambda}
		+ \text{h.~c.}
	\label{eq:l.ps.RKL}
	\;,
\end{align}
\end{subequations}
where the $\Gamma$ takes care of parity conservation.
We use
\begin{equation}
\Gamma =
  \begin{cases}
	\id & \text{for odd parity} \\
	\im \gamma^5 & \text{for even parity} \;,
  \end{cases}
  \notag
\end{equation}
and h.~c.~in Eqs.~\eqref{eq:lag.ps} denotes the \emph{hermitian conjugate}.

The pseudovector Lagrangians involve the derivative of the pion
wave function rather than the wave function itself. This introduces an additional mass
dimension, which is taken care of by a ``rescaling'' of the coupling constant.
It also ensures the matching of the on-shell behaviour the two types of Lagrangians.
The pseudovector Lagrangians are given by
\begin{subequations}\label{eq:lag.pv}
\begin{align}
	\mathcal{L}_{\pi N N_{1/2}^*}^{\text{PV}} & =
		- \frac{g_{\pi N N^*}}{m_{N^*} \pm m_{N}}
		\bar{\psi}_{N^*} \gamma^\mu \Gamma \partial_\mu
		( \vec{\tau} \cdot \vec{\phi}_{\pi} ) \psi_{N}
		+ \text{h.~c.}
	\label{eq:l.pv.piNR} \;, \\
	\mathcal{L}_{N_{1/2}^* K\Lambda}^{\text{PV}} & =
		- \frac{g_{N^* K \Lambda}}{m_{N^*} \pm m_{\Lambda}}
		\bar{\psi}_{N^*} \gamma^\mu \Gamma \partial_\mu \phi_{K} \psi_{\Lambda}
		+ \text{h.~c.}
	\label{eq:l.pv.RKL}
	\;,
\end{align}
\end{subequations}
where $\Gamma$ is given by
\begin{equation}
\Gamma =
  \begin{cases}
	\im & \text{for odd parity} \\
	\gamma^5 & \text{for even parity} \;,
  \end{cases}
  \notag
\end{equation}
and the upper and lower signs are used for even and odd parity resonances, respectively.

The spin-$3/2$ resonance Lagrangians are given by
\begin{subequations}\label{eq:lag.32}
\begin{align}
	\mathcal{L}_{\pi N N_{3/2}^*} & = \frac{g_{\pi N N^*}}{m_{\pi}}
		\bar{\psi}_{N^*}^\mu \partial_\mu (\vec{\tau} \cdot \vec{\phi}_{\pi})
		\psi_{N} + \text{h.~c.}
	\label{eq:l32.piNR} \;, \\
	\mathcal{L}_{N_{3/2}^* K \Lambda} & = \frac{g_{N^* K \Lambda}}{m_{K}}
		\bar{\psi}_{N^*}^\mu \partial_\mu \phi_{K}
		\psi_{\Lambda} + \text{h.~c.}
	\label{eq:l32.RKL}
	\;.
\end{align}
\end{subequations}
Such a form of the coupling was used for studying the
hypernuclear production also in other reactions~\cite{shy95,shy99}.
The values and signs of the various coupling constants have been taken from
Refs.~\cite{shy99,shy04} and are shown in table~\ref{tab:couplings}.
These parameters describe well the associated $K^+ \Lambda$ production
in proton-proton collisions within a similar resonance picture. All the
pion-resonance-kaon vertices that are of interest in this paper are involved
in this reaction. Thus the vertex parameters used by us inherently describe
the elementary process shown in Fig.~\ref{fig:Fig2}.
\begin{table}
\centering
\caption[T1]{Coupling constants for various vertices
used in the calculations.}\label{tab:couplings}
\begin{tabular}{c c}
\toprule
vertex & coupling constant ($g$)\\
\midrule
$N^*(1650)N\pi$        & 0.81  \\
$N^*(1650)\Lambda K^+$ & 0.76  \\
\midrule
$N^*(1710)N\pi$        & 1.04  \\
$N^*(1710)\Lambda K^+$ & 6.12  \\
\midrule
$N^*(1720)N\pi$        & 0.21  \\
$N^*(1720)\Lambda K^+$ & 0.87  \\
\bottomrule
\end{tabular}
\end{table}

\subsection{Resonance propagators}\label{sec:propagator}

The two interaction vertices of figure~\ref{fig:Fig2}
are connected by a resonance propagator.
For the spin-$1/2$ and spin-$3/2$ resonances the propagators are given by
\begin{equation}\label{eq:propspin12}
	\mathcal{D}_{1/2} = \im \frac{\gamma_\mu p^\mu + m}{p^2 - (m - \im\Gamma_{N^*}/2)^2}
\end{equation}
and
\begin{equation}\label{eq:propspin32}
	\mathcal{D}_{3/2}^{\mu\nu} = - \im
	\frac{\gamma_\lambda p^\lambda + m}{p^2 - (m - \im\Gamma_{N^*}/2)^2}
	P^{\mu\nu}
	\;,
\end{equation}
respectively. In Eq.~\ref{eq:propspin32} we have defined
\begin{equation}\label{eq:prop32proj}
P^{\mu\nu} =
\eta^{\mu\nu} - \frac{1}{3} \gamma^\mu \gamma^\nu - \frac{2}{3m^2} p^\mu p^\nu
	+ \frac{1}{3m} \left( p^\mu \gamma^\nu - p^\nu \gamma^\mu \right)
\;.
\end{equation}

$\Gamma_{N^*}$ in Eq.~\eqref{eq:propspin12} and Eq.~\eqref{eq:propspin32}
is the total width of the resonance.
It is introduced in the denominator term to account for the
finite life time of the resonances for decays into various
channels.
This is a function of the centre of mass momentum
of the decay channel, and it is taken to be the sum of the widths for pion
and rho decay (the other decay channels are considered only implicitly by
adding their branching ratios to that of the pion channel)~\cite{shy99}.
We have not introduced any correction in the resonance propagators to
account for the nuclear medium effects as no major change is expected in
our results due to
these effects. As pointed out in Refs.~\cite{lut03,eff97}, the medium
correction effects on the widths of the $s$- and $p$-wave resonances, which
make the dominant contribution to the cross sections investigated
here, are not substantial.
The reason for this is that resonances occur only as intermediate states
which implies an integration over their respective spectral distributions.

\section{Nuclear model} \label{chap:nucleus}

The spinors for the final bound hypernuclear state (corresponding to
momentum $p_\Lambda$) and for the intermediate nucleonic state
(corresponding to momenta $p_N$) are required to
perform numerical calculations of various amplitudes. We assume these
states to be of pure-single particle or single-hole configurations
with the core remaining inert.
In experimental measurements, however, core excited
states have also been detected (see, e.g.,~\cite{has06}). A covariant
description of the core polarisation can, in principle, be achieved
by following the method discussed, e.g., in Ref.~\cite{Ko:07}. This
procedure is somewhat tedious and is out of the scope of our present
study. Therefore, in this paper we concentrate on those transitions
which involve pure single-particle and single-hole states.

The spinors in momentum space are
obtained by the Fourier transformation of the corresponding coordinate space
spinors which are solutions of the Dirac equation with potential fields
consisting of an attractive scalar part ($V_\text{s}$) and a repulsive vector part
($V_\text{v}$) having a Woods--Saxon form. This choice appears justified as the
Dirac--Hartree--Fock calculations in Refs.~\cite{mil72,bro78} suggest that these
potentials tend to follow the nuclear shape. The same potential form has
also been used in the relativistic one-nucleon model~\cite{coo82a,coo82b} and
two-nucleon model calculations~\cite{shy95} of the $(p,\pi)$ reaction.

\subsection{Nucleon bound states}\label{sec:nuc.wav}

In our approach for describing the nucleon and hyperon wave functions,
we begin with the Dirac equation which is modified by introducing a
scalar and a time-like vector potential.
The modified Dirac equation with potentials is written as
\begin{equation}\label{eq:nw.pot}
\left(
	\im\gamma^\mu \partial_\mu - m
	- \gamma^0 V_{\text{v}} - V_{\text{s}}
\right) \psi = 0
\;.
\end{equation}

The solution, transformed to momentum space, can be written as in Ref.~\cite{shy06}
\begin{equation}\label{eq:mswf.wf}
\widehat{\psi}(p) = \delta(p_0-E)\left(
\begin{matrix}
	\hat{f}_{n, j}(k) \mathcal{Y}_{\ell s}^{j m_j}(\hat{p}) \\
	- \im \hat{g}_{n, j}(k) \mathcal{Y}_{\ell' s}^{j m_j}(\hat{p})
\end{matrix}
\right)
\;,
\end{equation}
where `$p$' is the four momentum vector of the particle.
`$\vec{p}$' represents the corresponding three momentum;
its magnitude $|\vec{p}|$ is denoted by $k$ and its
direction by $\hat{p}$.
Since we work in momentum space throughout this work,
we will drop the hats from the Fourier-transformed functions
$\psi$, $f$, and $g$.

The spin-spherical harmonics $\mathcal{Y}_{\ell s}^{j m_j}$ are given by
\begin{equation}\label{eq:nw.sphar1}
\mathcal{Y}_{\ell s}^{j m_j}(\hat{x}) = \sum_{m_\ell, m_s}
\langle \ell, m_{\ell}, s, m_s | j, m_j \rangle
Y_{\ell m_\ell}(\hat{x}) \chi_{s, m_s}
\;,
\end{equation}
where $\chi$ is the usual two-dimensional Pauli-spinor and $Y_{\ell m}$ are the
spherical harmonics of the first kind,
$\ell' = 2j - \ell$, and $m_\ell + m_s \stackrel{!}{=} m_j$,
and both are coupled to good total angular momentum.

The potentials in Eq.~\eqref{eq:nw.pot} have radial shapes of the
Woods--Saxon type.
The depths of the potential $V_{0\alpha}$, the radius $R_\alpha =
r_{0\alpha}A^{1/3}$, and the diffuseness $a_\alpha$
are treated as free parameters which are fixed by fitting
to the experimental data of the charge radius,
the nucleon separation energy, and the first diffraction minimum
of the charge form factor for the nucleon (see Ref.~\cite{Peters1998}).
Table~\ref{tab:pot.params} lists the potential parameters used for the
nuclear bound states.
\begin{table}
\centering
\caption{Potential parameters of the nuclear vector and scalar
potentials. For $^{12}$C and $^{40}$Ca, the potentials are fitted to the
experimental data on the charge radius, the nucleon separation energy,
and the first diffraction minimum of the charge form factor for the
nucleon (Ref.~\cite{Peters1998}). For $^{51}$V and $^{89}$Y, the
potential depths are fitted to the neutron separation energies with
radial parameters being the same as those of $^{40}$Ca.}
\label{tab:pot.params}
\begin{tabular}{ c c c c c c c }
\toprule
nucleus & $V_{\text{v}}\,$[MeV] & $r_{0\text{v}}\,$[fm]
& $a_{\text{v}}\,$[fm] & $V_{\text{s}}\,$[MeV]
& $r_{0\text{s}}\,$[fm] & $a_{\text{s}}\,$[fm] \\
\midrule
$^{12}$C & 385.7 & 1.056 & 0.427 & -470.4 & 1.056 & 0.447 \\
$^{40}$Ca & 348.1 & 1.149 & 0.476 & -424.5 & 1.149 & 0.506 \\
$^{51}$V & 309.7 & 1.149 & 0.476 & -382.3 & 1.149 & 0.506 \\
$^{89}$Y & 317.8 & 1.149 & 0.476 & -392.3 & 1.149 & 0.506 \\
\bottomrule
\end{tabular}
\end{table}

\subsection{Hyperon bound states}\label{sec:lam.wav}

The $\Lambda$, in contrast to a nucleon, can occupy any
bound state since it is not subject to the Pauli exclusion principle
with respect to the nucleons.
This property makes it an excellent probe for single bound states in a
nuclear potential.

In this case too, the spinors in the momentum space were obtained by
Fourier transforming the corresponding coordinate space spinors which
are solutions of the Dirac equation with potential fields consisting
of an attractive scalar part and a repulsive vector part having a
Woods-Saxon form.
With a fixed set of the geometry parameters (reduced radii
$r_\text{s}$ and $r_\text{v}$ and diffusenesses $a_\text{s}$ and $a_\text{v}$),
the depths of the
potentials were searched in order to reproduce the $\Lambda$ separation energy of
the particular state (the corresponding values are given in
table~\ref{tab:lpot.params} for $^{12}\!\!\!_{\Lambda}$C and
$^{51}\!\!\!_{\Lambda}$V, and in table~\ref{tab:lpot.params.y} for
$^{89}\!\!\!_{\Lambda}$Y).
We use the same geometry for the scalar and vector potentials. The
depths of potentials $V_\text{s}$ and $V_\text{v}$ were further constrained by
requiring that their ratios are equal to $-0.81$ as suggested in Ref.~\cite{ser86}.
Experimental inputs have been used for the
configurations and the single baryon binding energies of the states in
hypernuclei
$^{12}\!\!\!_\Lambda$C in Refs.~\cite{ban90,may81a,may81b},
$^{51}\!\!\!_\Lambda$V in Ref.~\cite{ban90}, and
$^{89}\!\!\!_\Lambda$Y in Ref.~\cite{ban90}.
However, the single baryon binding energies for states in
$^{40}\!\!\!_\Lambda$Ca hypernucleus were taken from the density
dependent relativistic hadron field (DDRH) theory predictions of
Refs.~\cite{kei00a,kei00b}, which reproduce the corresponding experimental binding
energies reasonably well. In this case, we have compared, for each state,
the spinors calculated by our well-depth search method with those
calculated within the DDRH theory (see Refs.~\cite{kei00a,kei00b}) and find an excellent
agreement between the two.
\begin{figure}
	\centering
	\includegraphics[width=0.50\textwidth]{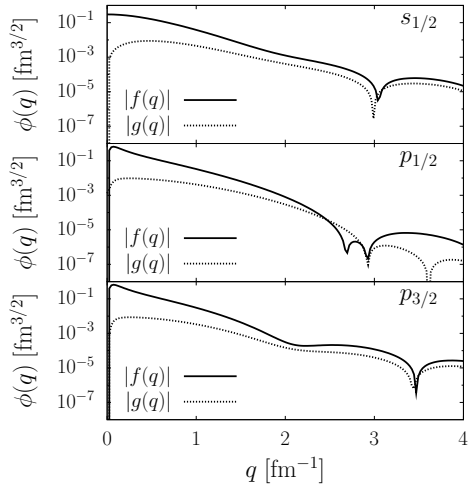}
	\caption{The $^{12}\!\!\!_{\Lambda}$C bound states in momentum space.}
	\label{fig:Fig3}
\end{figure}
\begin{table}
\centering
\caption{Potential parameters of the $\Lambda$ vector and scalar potentials of
the $^{12}\!\!\!_{\Lambda}$C and the $^{51}\!\!\!_{\Lambda}$V hypernuclei.
The $^{12}\!\!\!_{\Lambda}$C geometry is fixed to
$r_{0\text{v}} = 1.1486\,$fm, $a_{0\text{v}} = 0.3960\,$fm and
$r_{0\text{s}} = 1.1207\,$fm, $a_{0\text{s}} = 0.4764\,$fm.
The $^{51}\!\!\!_{\Lambda}$V geometry is fixed to
$r_{0\text{v}} = 0.9827\,$fm, $a_{0\text{v}} = 0.5779\,$fm and
$r_{0\text{s}} = 0.9825\,$fm, $a_{0\text{s}} = 0.6064\,$fm.}
\label{tab:lpot.params}
\begin{tabular}{ c c c c }
\toprule
orbital & $E_{\text{bind}}\,$[MeV] & $V_{0\text{v}}\,$[MeV] & $V_{0\text{s}}\,$[MeV] \\
\midrule
$^{12}\!\!\!_{\Lambda}$C($s_{1/2}$) & $10.79\pm 0.11$ & 171.5230 & -211.7654 \\
$^{12}\!\!\!_{\Lambda}$C($p_{3/2}$) & $0.10\pm 0.04$  & 171.5230 & -211.7654 \\
\midrule
$^{51}\!\!\!_{\Lambda}$V($s_{1/2}$) & $19.75$ & 151.2006 & -186.6674 \\
$^{51}\!\!\!_{\Lambda}$V($p_{3/2}$) & $11.75$ & 171.2202 & -211.3830 \\
$^{51}\!\!\!_{\Lambda}$V($d_{5/2}$) &  $3.75$ & 197.5942 & -243.9435 \\
\bottomrule
\end{tabular}
\end{table}
\begin{table}
\centering
\caption{Potential parameters of the $\Lambda$ vector and scalar potentials of
the $^{89}\!\!\!_{\Lambda}$Y hypernucleus. The geometry is fixed to
$r_{0\text{v}} = 0.9827\,$fm, $a_{0\text{v}} = 0.5779\,$fm and
$r_{0\text{s}} = 0.9825\,$fm, $a_{0\text{s}} = 0.6064\,$fm.}
\label{tab:lpot.params.y}
\begin{tabular}{ c c c c }
\toprule
orbital & $E_{\text{bind}}\,$[MeV] & $V_{0\text{v}}\,$[MeV] & $V_{0\text{s}}\,$[MeV] \\
\midrule
$^{89}\!\!\!_{\Lambda}$Y($s_{1/2}$) & $23.1$ & 149.3054 & -184.3276 \\
$^{89}\!\!\!_{\Lambda}$Y($p_{1/2}$) & $16.5$ & 171.1863 & -211.3411 \\
$^{89}\!\!\!_{\Lambda}$Y($p_{3/2}$) & $16.5$ & 165.3949 & -204.1913 \\
$^{89}\!\!\!_{\Lambda}$Y($d_{3/2}$) & $10.0$ & 208.0682 & -256.8744 \\
$^{89}\!\!\!_{\Lambda}$Y($d_{5/2}$) & $10.0$ & 190.7247 & -235.4626 \\
$^{89}\!\!\!_{\Lambda}$Y($f_{5/2}$) &  $2.3$ & 253.8945 & -313.454  \\
$^{89}\!\!\!_{\Lambda}$Y($f_{7/2}$) &  $2.3$ & 214.0216 & -264.2242 \\
\bottomrule
\end{tabular}
\end{table}

\begin{figure}
	\centering
	\includegraphics[width=0.50\textwidth]{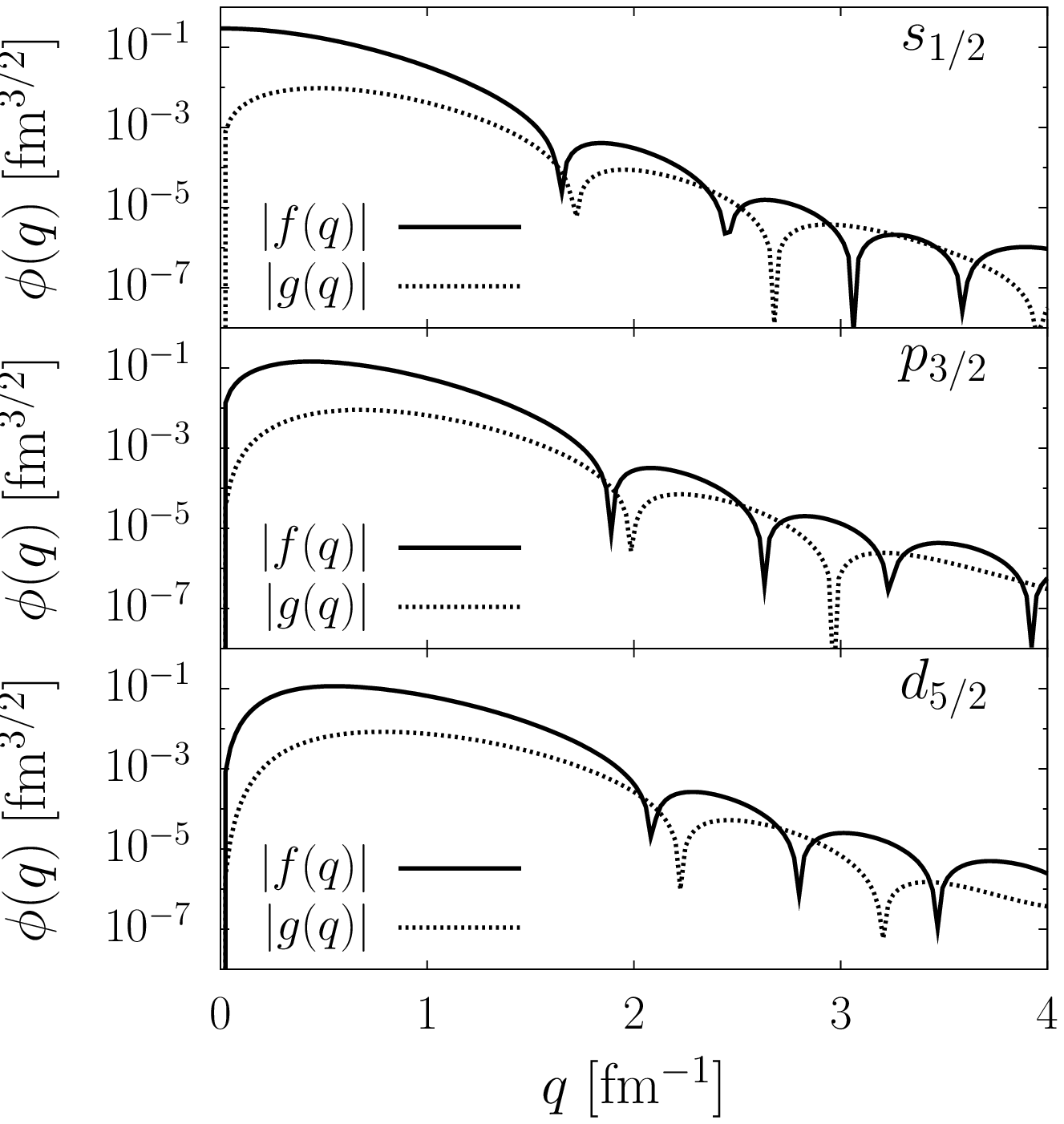}
	\caption{The $^{51}\!\!\!_{\Lambda}$V bound states in momentum space.}
	\label{fig:Fig4}
\end{figure}

\begin{figure}
	\centering
	\includegraphics[width=0.60\textwidth]{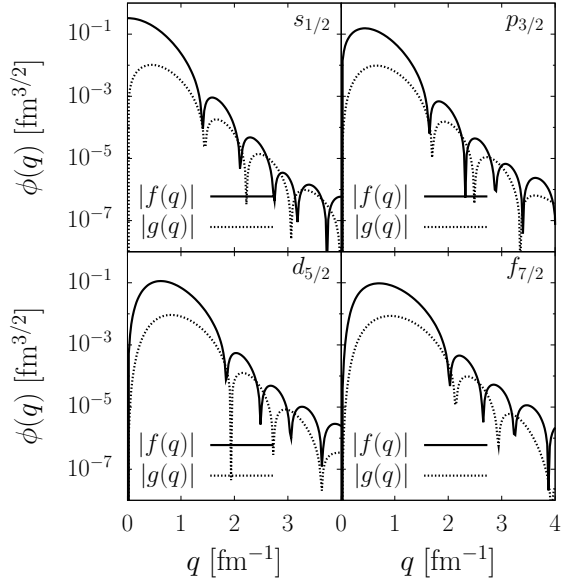}
	\caption{The $^{89}\!\!\!_{\Lambda}$Y bound states in momentum space.}
	\label{fig:Fig5}
\end{figure}

The upper and lower components of the Dirac spinors for the
$\Lambda$ bound states in momentum space are shown in
figures~\ref{fig:Fig3}--\ref{fig:Fig5} for
$^{12}\!\!\!_{\Lambda}$C, $^{51}\!\!\!_{\Lambda}$V, and
$^{89}\!\!\!_{\Lambda}$Y, respectively.
We note that in each case, only for momenta $q<2\,$fm$^{-1}$ is the
lower component of the spinor substantially smaller than the upper
component.  In the region of momentum transfer pertinent to exclusive
kaon production in pion-nucleus collisions, the lower components of
the spinors are not negligible as compared to the upper component which
clearly demonstrates that a fully relativistic approach is better for an
accurate description of this reaction.

\section{Calculation of the cross section}
\label{chap:reaction}

We use the subscripts $\pi$, $K$, $A$, and $B$ to denote the quantities
of the incoming pion, the outgoing kaon, the target nucleus, and the
final hypernucleus, respectively.
The differential cross section for the $(\pi^+, K^+)$ reaction
is given by
\begin{equation}\label{eq:xs.2}
	\ud\sigma = \frac{1}{(2\pi)^2}
	\, \frac{\ud^3 p_{K}}{2E_{K}}
	\, \frac{\ud^3 p_{B}}{2E_{B}}
	\, \frac{m_{A} m_{B}}{|\vec{p}_{\pi}|\sqrt{s}}
	\, \left\lvert\sum_{R_i} \mathcal{M}_{R_i}\right\rvert^2
	\delta^{(4)}\left( p_{\pi}+p_{A} - (p_{K} + p_{B}) \right)
	\;.
\end{equation}
The summation is carried out over initial ($m_i$) and final ($m_f$)
spin states; $\sum_{R_i}$ indicates the summation over all three
resonances.

After having established the effective Lagrangians and the coupling
constants, one can  write down, by following the well known
Feynman rules, the amplitudes for the graph shown in figure~\ref{fig:Fig2}.
The isospin part is treated separately which gives rise to a constant
factor for each graph.
Thus, the matrix element for our process is given by
\begin{multline}\label{eq:tmatrix}
\mathcal{M} = \int\! \frac{\ud^4 k_{N}}{(2\pi)^4}
	\int\! \frac{\ud^4 k_{\Lambda}}{(2\pi)^4}
	\int\! \frac{\ud^4 p}{(2\pi)^4}
	\, \phi_{K}^*(p - k_{\Lambda})
	\bar{\psi}_{\Lambda}(k_{\Lambda}) \Gamma_\alpha \\
	\times \im \frac{\gamma_\mu p^\mu + m_{N^*}}{p^2 - ( m_{N^*}^2 -
	\im\Gamma_{N^*}/2 )^2 }
	\Gamma_\beta \phi_{\pi}( p - k_{N} )
	\psi_{N}(k_{N})
	\;.
\end{multline}
In Eq.~\eqref{eq:tmatrix}, the factors $\Gamma_\alpha$ and
$\Gamma_\beta$ are given by the interaction Lagrangians from
Eqs.~\eqref{eq:lag.ps}--\eqref{eq:lag.32}.
The $\phi$s denotes the meson wave functions, and the $\psi$s are the
solutions of the in-medium single-particle Dirac equations
given in sections~\ref{sec:nuc.wav} and~\ref{sec:lam.wav}.

The incident pion and outgoing kaon fields are given by
\begin{align}
\phi_{\pi}^{(+)}(p_\pi^\prime) & = \delta(p^\prime_{\pi 0} - E_\pi)
	\sum_{\ell_\pi m_\pi} (-1)^{\ell_\pi} Y^*_{\ell_\pi m_\pi} (\hat{p}_\pi)
	Y_{\ell_\pi m_\pi} (\hat{p}^\prime_\pi) \notag \\
& \qquad \times  f_{\ell_\pi}(k^\prime_\pi, k_\pi)
\;, \label{eq:mesonwf.pi} \\
\phi_{K}^{(-)*}(p_K^\prime) & = \delta(p^\prime_{K0} - E_K)
	\sum_{\ell_K m_K} (-1)^{\ell_K} Y_{\ell_K m_K} (\hat{p}_K)
	Y_{\ell_K m_K}^*(\hat{p}^\prime_K) \notag \\
& \qquad \times  f_{\ell_K}(k^\prime_K, k_K)
\;, \label{eq:mesonwf.ka}
\end{align}
where $E_\pi$ and $E_K$ represents the energies of the incident pion and
outgoing kaon, respectively; $p_K$ and $p_\pi$ denote the meson on-shell
momenta.
The functions $f_\ell$ are given by
\begin{equation}\label{eq:meson.fl}
	f_\ell(k') = \frac{1}{2\pi^2} \int_{0}^{\infty}
	j_\ell(k' r) f_\ell^C(r) r^2 \ud r
	\;,
\end{equation}
where the wave function $f^C_\ell$ is the coordinate space solution of
the Klein--Gordon equation with a meson-nucleus optical potential (see,
e.g., Refs.~\cite{tab77a,tab77b}).

It should be mentioned here that due to the oscillatory nature
of the wave functions in the asymptotic region, the
integrals involved in Eq.~\eqref{eq:mesonwf.pi} and in Eq.~\eqref{eq:mesonwf.ka},
and in a similar fashion the radial components of the bound state wave
functions in Eq.~\eqref{eq:mswf.wf}, converge poorly.
Such integrals can, however, be calculated very accurately by using a
contour integration method as in Ref.~\cite{dav88}.

In the plane wave approximation, the wave functions
$\phi_\pi^{(+)}$ and $\phi_{K}^{(-)*}$ are given by
\begin{align}
\phi_{\pi}^{(+)}(p_\pi^\prime) & = \delta^4(p^\prime_\pi - p_\pi) \;,
\label{eq:pwwf.pi} \\
\phi_{K}^{(-)*}(p_K^\prime) & = \delta^4(p^\prime_K - p_K) \;.
\label{eq:pwwf.ka}
\end{align}
Thus, the integration over $k_N$ and $k_\Lambda$ become redundant.
This not only reduces the dimensionality of the integrations
by a factor of eight, but also removes the requirement of partial wave
summations altogether.

For a fully dynamical description of the production of hypernuclei,
however, we have to consider the interactions of the incoming and
outgoing particles with the nucleus.
In this exploratory study, however, we would like to
make calculations for very many cases involving a variety of targets and
beam energies in order to understand the basic mechanism of this reaction
and to get the relative estimates of the order of magnitudes of various cross
sections within a covariant model.
Therefore, we restrict ourselves
to the plane wave treatment of the scattering states in the
initial and final channels.
The weak mutual interaction (kaon-nucleus)
in the final channel and higher energies of the projectile in
the initial channels do provide support to this choice. Indeed,
it has been shown in Ref.~\cite{mot88} that in the non-relativistic
impulse approximation calculations the pattern of the spectra
look similar in the PW and DW calculations---only the absolute
magnitudes of the cross sections are affected by the initial and
final channel meson-nucleus interactions.
Indeed, distortion effects from an optical potential $U_{\text{opt}}$
should contribute in the leading order by
$\mathcal{O}(|U_{\text{opt}}/{E}|)$, which is well below unity at the energies
around $1.0\,$GeV considered in this work.

Nevertheless, we have made an estimate of the influence of the initial state
interaction (pion-nucleus channel) on the magnitudes of the ($\pi^+,K^+$)
cross sections within an eikonal approximation.

\subsection{The eikonal approximation for the pion-nucleus interaction} \label{sec:mes.pie}

The eikonal approximation (see Refs.~\cite{Wallace1971, JO1975}) has been quite
successful in describing the scattering of pions on nuclei at higher incident
energies~\cite{Germond1979, Cha1996}. In this approximation, the relative motion
wave function is given by (see for example Ref.~\cite{JO1975})
\begin{equation}\label{eq:eik.wf1}
	\phi(\vec{x}) = \exp\left\{ \im \vec{k}\cdot\vec{x}
		- \frac{\im}{v} \int_{-\infty}^{z} V(\vec{b},z') \,\ud z' \right\}
\;,
\end{equation}
where $\vec{k}$ is the incident (asymptotic) momentum of the particle,
$v = |\vec{k}|/m$ is the magnitude of the incident velocity, and $\vec{b}$
is the (two-dimensional) impact vector in cylindrical coordinates.
The optical potential is, in general, a complex function $V = U - \im W$ which
results in a phase factor from the real part $U$ and an amplitude
reduction from the imaginary part $W$.

Since at higher energies several effects are suppressed, a simple form
for the optical potential can be used.
For example, the $t\varrho$-approximation (see, e.g., Ref.~\cite{JO1975})
relates the potential to the
free-space single-particle scattering amplitudes (or to the total cross
section) and the density.
The optical potential in this case is given by
\begin{align}\label{eq:eik.trho}
	V_{\text{opt}}(\vec{x}) &= - \frac{4\pi}{2E_{\text{lab}}}
	\left[ f_{mp} \varrho_{p}(\vec{x})
	+ f_{mn} \varrho_{n}(\vec{x}) \right] \notag \\
	&= - \frac{k}{2E_{\text{lab}}}
	\left[ \im\sigma_{mp}^{\text{tot}}(1-\im\gamma_{mp}) \frac{Z}{A} +
	\im\sigma_{mn}^{\text{tot}}(1-\im\gamma_{mn}) \frac{N}{A} \right]
	\varrho(\vec{x})
	\;,
\end{align}
where $f_{mp}$ and $f_{mn}$ are the elementary
free-space meson-proton and meson-neutron scattering amplitudes, respectively.
Applying the optical theorem, they can be substituted by the
total cross sections $\sigma_{mp}^{\text{tot}}$ and
$\sigma_{mn}^{\text{tot}}$.
The respective ratios of the real to the imaginary part of the scattering
amplitudes are denoted by $\gamma_{mp} := \Im f_{mp}/\Re f_{mp}$
and $\gamma_{mn} := \Im f_{mn}/\Re f_{mn}$.
We have also separated the neutron and proton contributions to the potential, whereby $Z$ is
the proton number, $N$ is the neutron number, and $A$ is the total number of nucleons.

For spherically symmetric nuclei the density, and hence the potential, depends only on the
magnitude of $\vec{x}$, $r := |\vec{x}|$.
In these cases we can integrate Eq.~\eqref{eq:eik.wf1}
by rewriting the argument of the potential as
\begin{equation}\label{eq:eik.potbz}
	V(r) = V(\vec{b},z) = V\left(\sqrt{|\vec{b}|^2 + z^2}\right)
	\;.
\end{equation}

We approximate the density by relatively simple parameterisations which
are easier and faster to compute numerically. For light nuclei ($A\leq 16$),
a modified Gaussian shape has been used,
\begin{equation}\label{eq:eik.gauss}
	\varrho_{\text{G}}(r) = \frac{1}{(\sqrt{\pi}R_{\text{G}})^3} \left[
	4 + \frac{2(A-4)}{3} \frac{r^2}{R_{\text{G}}^2} \right]
	\ex^{- \frac{r^2}{R_{\text{G}}^2}}
	\;,
\end{equation}
with $R_{\text{G}}$ being the radial parameter.
For heavier nuclei ($A > 16$), a Woods--Saxon shaped density with the radial
parameter $R$ and the \emph{diffuseness} parameter $a$ has been used,
\begin{equation}\label{eq:eik.wsden}
	\varrho_{\text{WS}}(r) = \frac{\varrho_0}{ 1 + \exp{ \{\frac{r-R}{a}\} } }
\end{equation}
The densities are normalised to the total nucleon number, such that
\begin{equation}\label{eq:eik.densnorm}
	\int_{\mathbb{R}^3} \varrho(\vec{x}) \,\ud^3 x = A
	\;,
\end{equation}
which is already fulfilled in the Gaussian case.
In the Woods--Saxon case this determines $\varrho_0$, which is
given by Refs.~\cite{Satchler1980, Satchler1983}
\begin{equation}\label{eq:wsrho0}
	\varrho_0 = \frac{3A}{4\pi R^3} \frac{1}{1 + (\frac{\pi a}{R})^2}
	\;.
\end{equation}

We fit the parameters $R_{\text{G}}$, $R$ and $a$ to the radial densities.
The fitted values for both density approximations for $^{12}\!\!\!_{\Lambda}$C
and $^{40}\!\!\!_\Lambda$Ca are given in table~\ref{tab:eik.trho}.
\begin{table}
\centering
\caption{Density parameters for the eikonal approximation for
	$^{12}\!\!\!_{\Lambda}$C and $^{40}\!\!\!_\Lambda$Ca,
	fitted to the nucleon wave functions and the elastic
	scattering cross sections.}\label{tab:eik.trho}
\begin{tabular}{ c c c c c c }
	\toprule
	& $A$ & $Z$ & {Woods--Saxon} & { Gauss } \\
	& & & $R\,$[fm] & $a\,$[fm] & $R_{\text{G}}\,$[fm] \\
	\midrule
	$^{12}\!\!\!_{\Lambda}$C & 12 & 6 & 2.24 & 0.46 & 1.6 \\
	$^{40}\!\!\!_\Lambda$Ca & 40 & 20& 3.49 & 0.547 & 2.08 \\
	\bottomrule
\end{tabular}
\end{table}

\section{Results and discussion}
\label{chap:results}

\subsection{Cross sections for the A$(\pi^+,K^+)_\Lambda$A reaction}
\label{chap:res.pw}

\begin{figure}
\centering
\begin{tabular}{cc}
\includegraphics[scale=0.42]{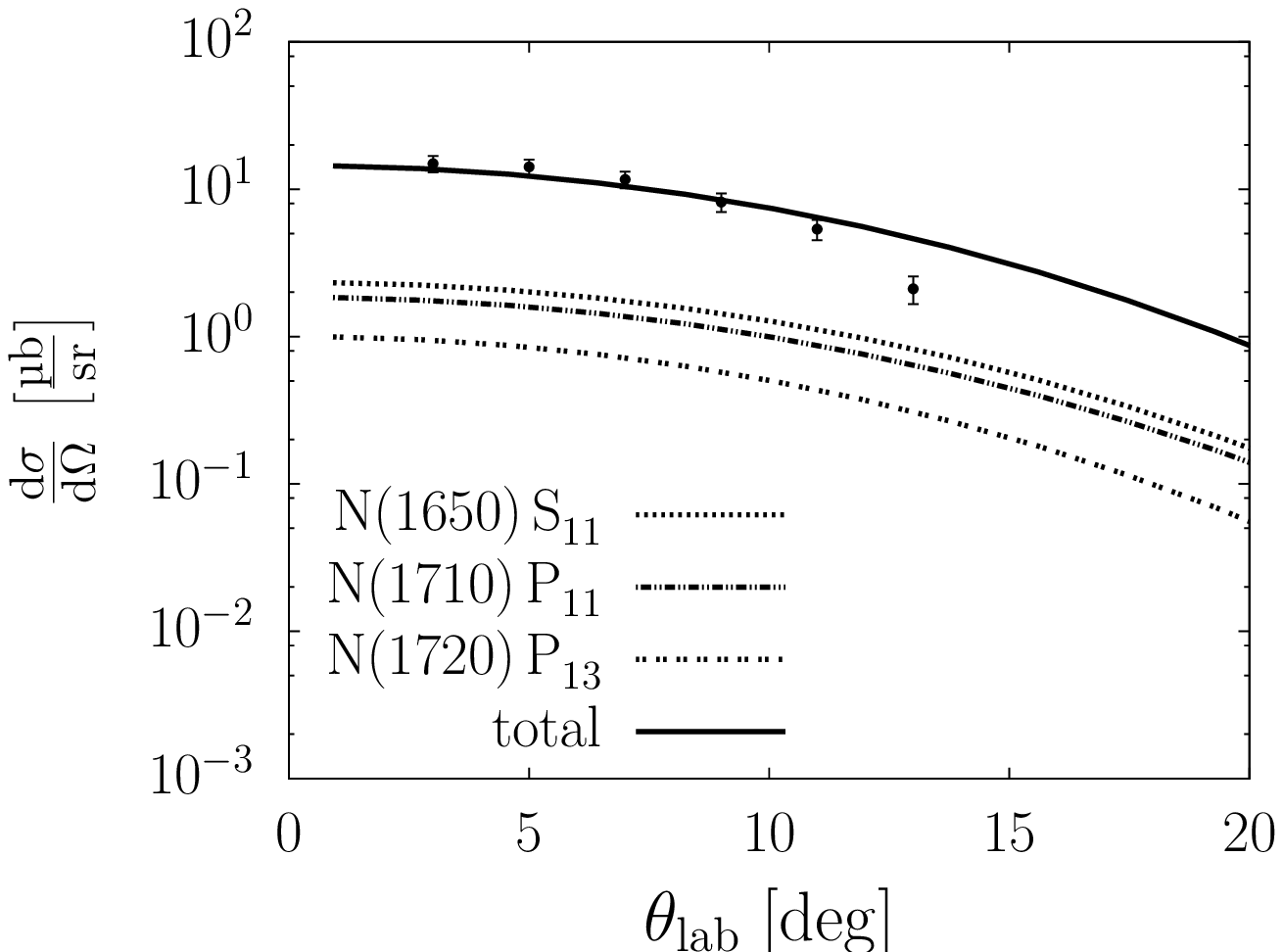} &
\includegraphics[scale=0.42]{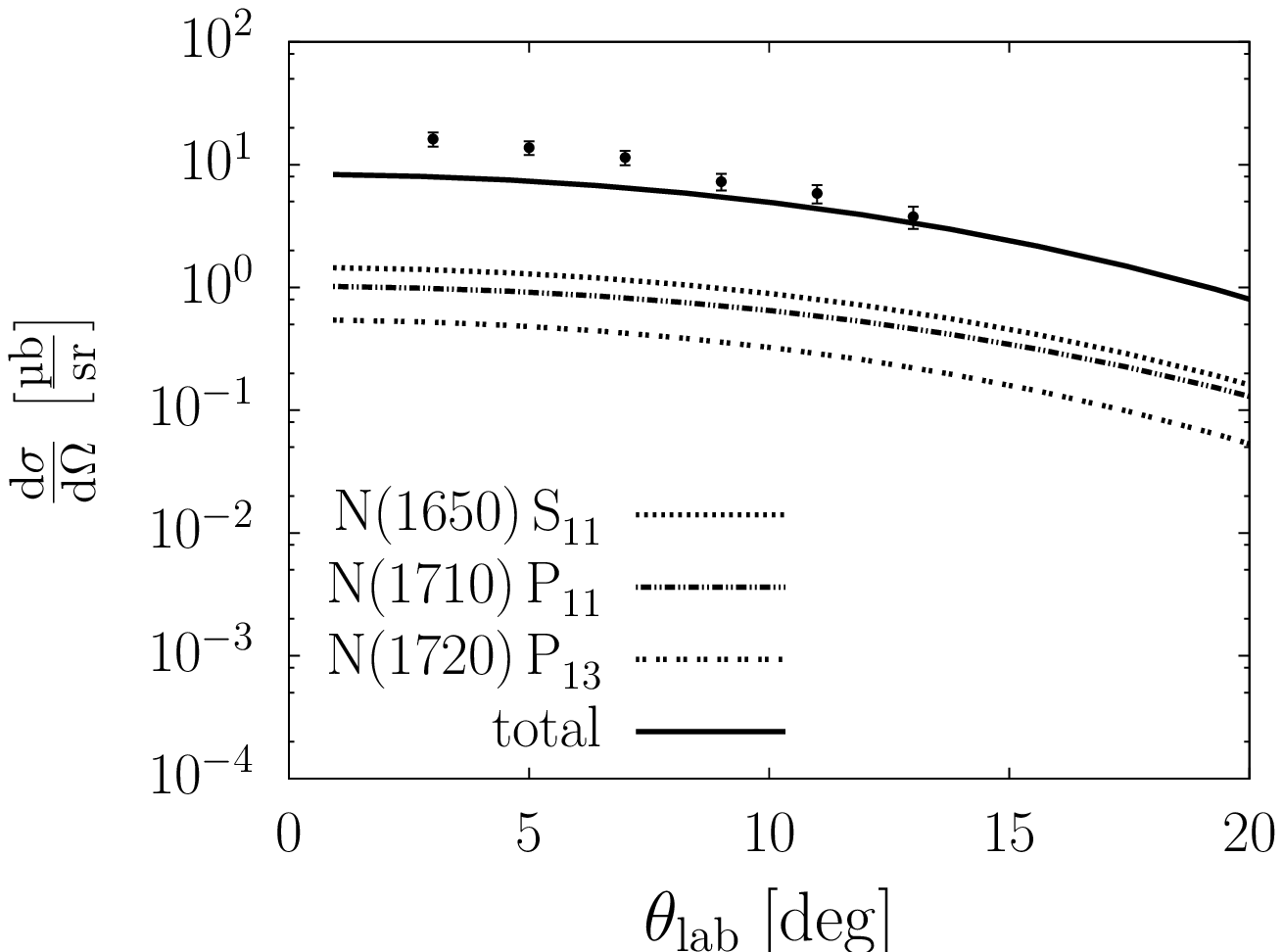}
\end{tabular}
\caption{The differential cross section for the
$\pi^+ + ^{12}$C $\rightarrow K^+ + ^{12}\!\!\!\!\!_\Lambda$C
reaction leading to states of the hypernucleus $^{12}\!\!\!_\Lambda$C
with $[(p_{3/2}^{-1})_N,(s_{1/2})_{\Lambda}]$ (left panel) and
$[(p_{3/2}^{-1})_N,(p_{3/2})_{\Lambda}]$ (right panel)
configurations. The lab momentum of the incoming pion ($p_{\text{lab}}$)
is $1.05\,$GeV.}
\label{fig:Fig6}
\end{figure}

First, we present our results for the
$^{12}$C$(\pi^+,K^+)^{12}\!\!\!_{\Lambda}$C reaction at an
incident pion lab momentum ($p_{\text{lab}}$) of $1.05\,$GeV.
In the left panel of Fig.~\ref{fig:Fig6}, we show a comparison of our
calculations with the experimental data (taken from Ref.~\cite{hot01b})
for the $K^+$ angular distribution of the
$^{12}$C$(\pi^+,K^+)^{12}\!\!\!_{\Lambda}$C reaction
where the hypernuclear state has the $[(p_{3/2}^{-1})_N,(s_{1/2})_{\Lambda}]$
configuration. The leading contribution to this transition comes from
the $1^-$ ground state of $^{12}\!\!\!_{\Lambda}$C
corresponding to the first peak seen in the hypernucleus spectrum.
Also shown in this figure are the contributions of the individual baryon
resonances. We note that while the contributions of the $N^*(1650)$ and
$N^*(1710)$ resonances are almost identical, those of the $N^*(1720)$ state is
weaker by factors of $3$--$5$. The dominant contribution of the two spin-$1/2$
resonances was also noted in case of the hypernuclear production
reactions studied via $(\pi^+,K^+)$ and $(\gamma,K^+)$ reactions in
Refs.~\cite{shy04,shy09a} within a model similar to that employed in
this paper.
We also note that interference effects of the resonances are important
as their individual contributions do not sum up to the total cross
sections shown by the solid lines.
We see that the agreement between the calculations and the experimental
data is better at forward angles while at higher angles the calculation
overestimate the data. This trend was also noted in the non-relativistic
distorted wave impulse approximation (DWIA) calculations as shown in Ref.~\cite{hot01b}.
This indicates that at larger momentum transfers the simple no-core
excitation picture may not be adequate. Indeed, in Ref.~\cite{has06} it is
shown that agreement of the DWIA calculations with the data improves if
contributions from the core excited states are also included in the
calculated cross sections. There could also be a weak contribution (about
$10\%$) from the $2^-$ member of the multiplet.

The next $\Lambda$-bound state in $^{12}\!\!\!_{\Lambda}$C has the
configuration $[(p_{3/2}^{-1})_N,(p)_{\Lambda}]$ which according to
Ref.~\cite{hot01b} is just bound by about $0.1\,$MeV. In the right
panel of Fig.~\ref{fig:Fig6}, we compare the calculated differential
cross section for this transition with the data (set 5 of Ref.~\cite{hot01b}).
The theoretical results correspond to the natural parity
$2^+$ state of the hypernucleus. In this case our calculations
underestimate the data at very forward angles. It should, however,
be noted that the data could also have contributions from other members of
the $[(p_{3/2}^{-1})_N,(p)_{\Lambda}]$ configuration and also from the core
excited configurations. Nevertheless, the DWIA calculations, which
include contributions from such configurations, overestimate the data
for this transition~\cite{hot01b} in the entire angular range.

\begin{figure}
\centering
\includegraphics[width=0.80\textwidth]{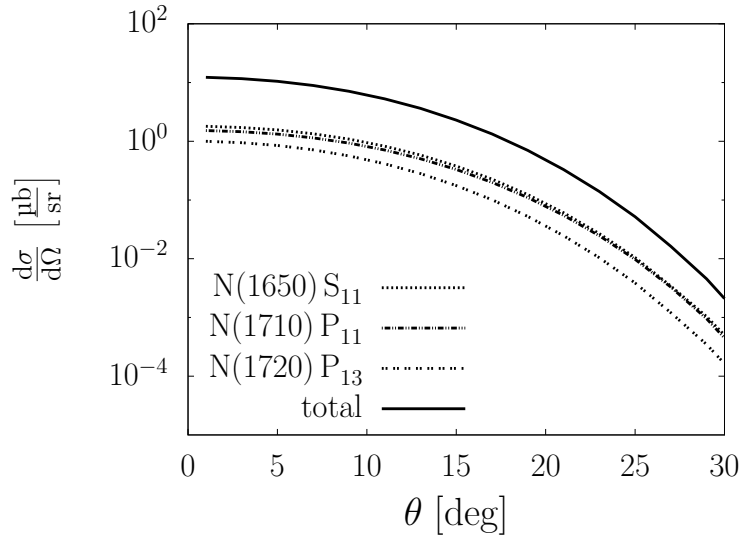}
\caption{The differential cross section for
$\pi^+ + ^{40}$Ca $\rightarrow K^+ + ^{40}\!\!\!_\Lambda$Ca
at a pion incoming momentum of $p_{\text{lab}} = 1.05\,$GeV
as a function of the kaon angle $\theta$.}
\label{fig:Fig7}
\end{figure}

In Fig.~\ref{fig:Fig7} we show the contributions of
various baryonic resonances to the angular distribution
of the $\pi^+ + ^{40}$Ca $\rightarrow K^+ + ^{40}\!\!\!_\Lambda$Ca
reaction corresponding to the $2^+$ state of $^{40}\!\!\!_\Lambda$Ca
having a $[(d_{3/2}^{-1})_N,(s_{1/2})_\Lambda]$ configuration. The
corresponding binding energy was taken to be $20\,$MeV which is
consistent with the experimental value reported in Ref.~\cite{chr88}.
We note that like the $^{12}$C, case the two spin-$1/2$ resonances
contribute almost equally to the cross sections and both are
larger than the contribution of the spin-$3/2$ one. Detailed experimental
data are not available for this case.  The single point
reported in Ref.~\cite{chr88} corresponds to the differential
cross section at zero angle which is ${\sim}9\,$mb/sr. Our result
for this case is close to this value.

\begin{figure}
\centering
	\includegraphics[width=0.80\textwidth]{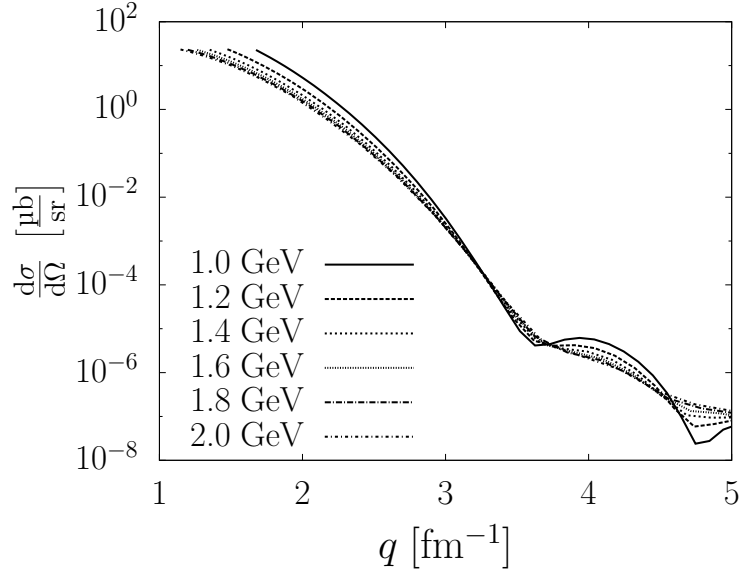}
	\caption{The differential cross section for the same reaction as in
	the left panel of Fig.~\ref{fig:Fig6}
	for some selected values of the incoming pion momentum as a function of the momentum
	transfer $q$.}
\label{fig:Fig8}
\end{figure}

In Fig.~\ref{fig:Fig8} we show the differential cross section
for the same reaction as that in the left panel of Fig.~\ref{fig:Fig6}
as function of the magnitude of the momentum transfer to the nucleus
($\vec{q} = \vec{p}_{\pi} - \vec{p}_{K}$) for several beam energies.
We see that shapes of the differential cross sections are qualitatively
different from each other at different beam energies. However, their
absolute magnitudes differ at forward angles which is more
prominent as the beam energy is increased from $1.0\,$GeV to $1.4\,$GeV. However,
for beam energies $> 1.4\,$GeV this difference is small. At larger momentum
transfers the cross sections are too small to be amenable to measurements.
Thus to get the larger cross section (i.e. counting rates) forward angles
experiments (where $q$ is smaller) with beam energies around $1\,$GeV
appear to be favourable.

\begin{figure}
\centering
\begin{tabular}{cc}
\includegraphics[scale=0.64]{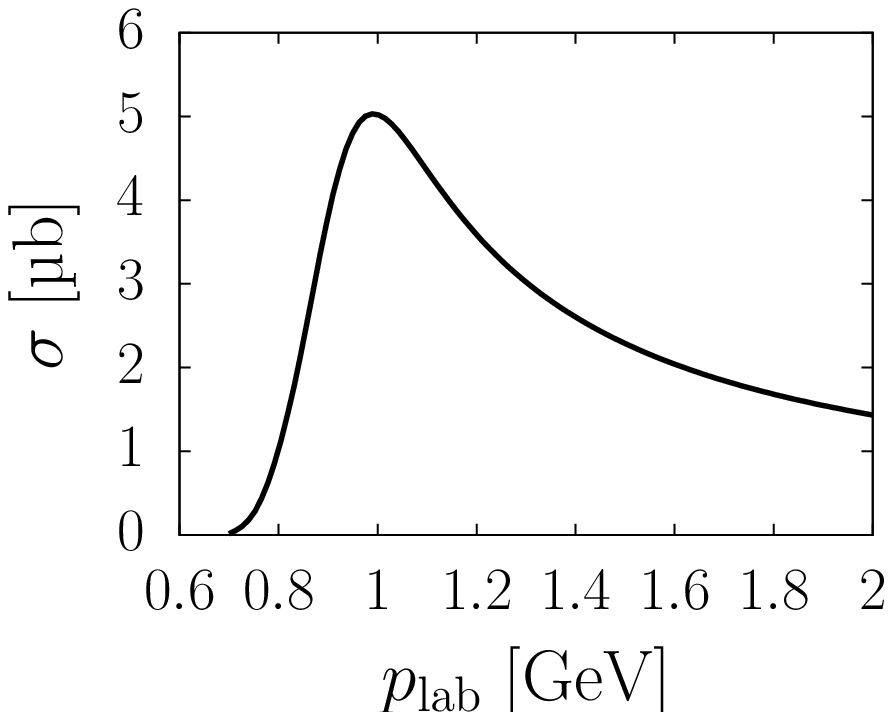} &
\includegraphics[scale=0.64]{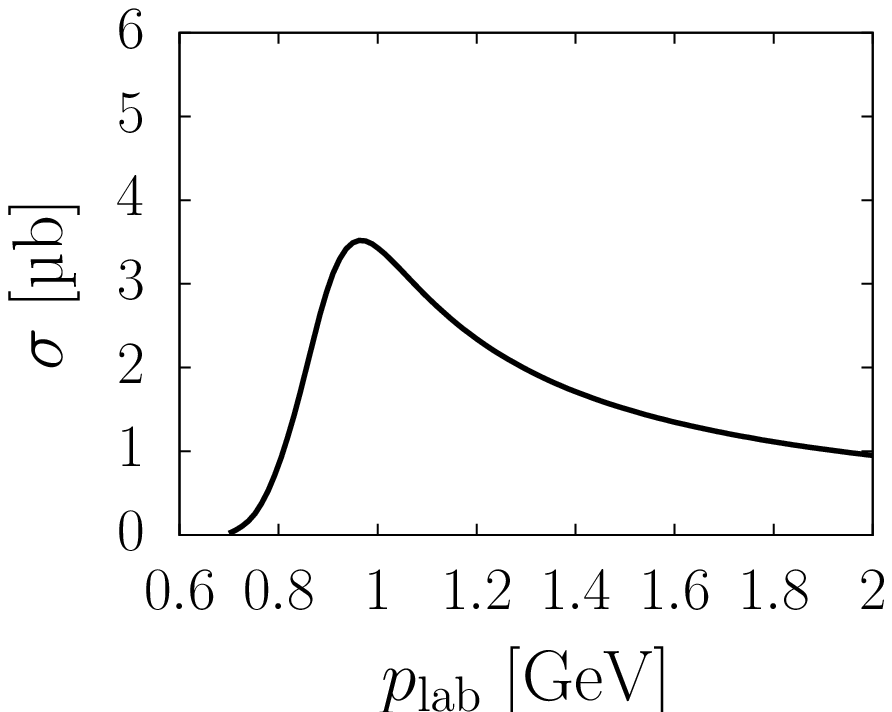}
\end{tabular}
\caption{The total cross section for the
$[(p_{3/2}^{-1})_N, (s_{1/2})_\Lambda]$ transition in the
$\pi^+ + ^{12}$C $\rightarrow K^+ + ^{12}\!\!\!\!\!_\Lambda$C reaction (left panel),
and the $[(d_{3/2}^{-1})_N, (s_{1/2})_\Lambda]$ transition in the
$\pi^+ + ^{40}$Ca $\rightarrow K^+ + ^{40}\!\!\!\!\!_\Lambda$Ca reaction (right panel)
as a function of the incident pion momentum $p_{\text{lab}}$.}
\label{fig:Fig9}
\end{figure}

In Fig.~\ref{fig:Fig9} we show the dependence of the total cross section
on $p_{\text{lab}}$ for the reactions studied in
Figs.~\ref{fig:Fig6} and~\ref{fig:Fig7}. We note that in both the cases,
as beam energy increases beyond the threshold, the cross sections first
increases rapidly and after peaking around a given value starts decreasing
slowly. The beam momentum where the cross section peaks is around $1.0\,$GeV.
This observation was also made in the DWIA study of this reaction in
Ref.~\cite{dov80}. This further highlights the point made in Fig.~\ref{fig:Fig8}.

\begin{figure}
\centering
	\includegraphics[width=0.50\textwidth]{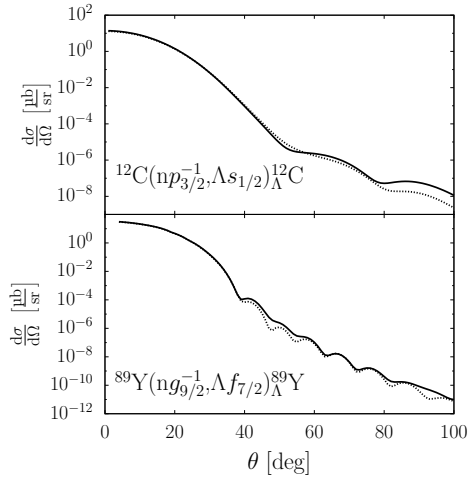}
	\caption{The differential cross section for the same reaction as in
	the left panel of Fig.~\ref{fig:Fig6} and for the
	$[(g_{9/2}^{-1})_N,(f_{7/2})_{\Lambda}]$ configuration in the
	$\pi^+ + ^{89}$Y $\rightarrow K^+ + ^{89}\!\!\!\!\!_\Lambda$Y
	reaction. Shown are the results including the full relativistic
	spinors (solid lines) and without the lower components (dotted lines)
	as a function of the kaon angle $\theta$.}
\label{fig:Fig12}
\end{figure}

Next we examine the role of the bound state relativistic effects on the
cross sections. We would like to point out that since our calculations
have been performed in momentum space, there is no need to introduce any
local approximation to the propagators appearing in the production
amplitudes---most of the non-relativistic calculations necessarily make such
approximation. In Fig.~\ref{fig:Fig12}, we show the role of the lower component of the
Dirac bound state wave functions ($g$) on the angular distributions of the
$(\pi^+,K^+)$ reaction on the $^{12}$C and $^{89}$Y targets for the transitions
as depicted in this figure. We see that effect of $g$ is more pronounced at
larger angles. It should, however, be noted that the results with upper
components only can not be directly equated with those of the conventional
approaches that employ the Sch\"odinger equation to describe the bound state
wave functions. One requires the matrix elements and particularly the operator
that is given in terms of the Dirac matrices to undergo a non-relativistic
reduction. Such a reduction has been carried out in
Refs~\cite{Bennhold:1987cb,Bennhold:1989bw} where it is shown that for the
hypernuclear production by the $(\gamma,K^+)$ reaction, the difference
between the relativistic and non-relativistic cross sections could be about
20$\%$ even in the forward directions.

\subsubsection{Spectral distributions for $^{51}\!\!\!_{\Lambda}$V
and $^{89}\!\!\!_{\Lambda}$Y hypernuclei}\label{sec:res.v51.y89}

\begin{figure}
\centering
\includegraphics[width=0.80\textwidth]{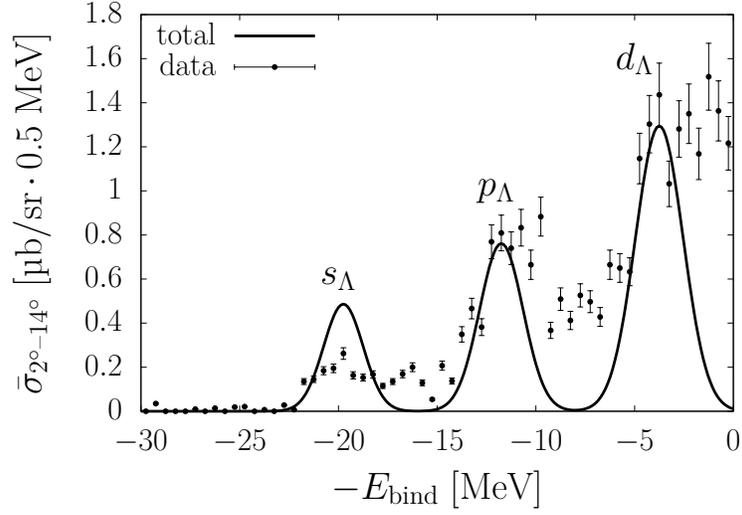}
\caption{The spectral distribution for the
$^{51}\!\!\!_{\Lambda}$V hypernucleus.}
\label{fig:Fig10}
\end{figure}

\begin{figure}
\centering
\includegraphics[width=0.80\textwidth]{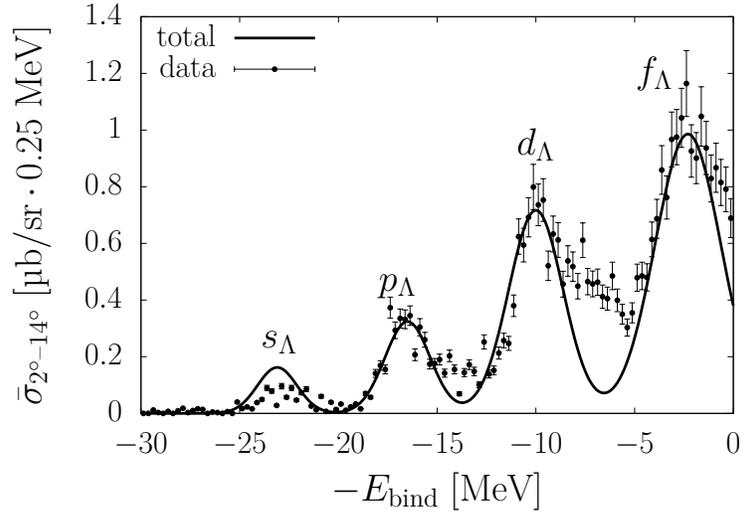}
\caption{The total averaged cross section for the
$^{89}\!\!\!_{\Lambda}$Y hypernucleus.}
\label{fig:Fig11}
\end{figure}

In Figs.~\ref{fig:Fig10} and~\ref{fig:Fig11}, we show the comparison of the calculated spectral
distributions for the $^{51}\!\!\!_{\Lambda}$V and $^{89}\!\!\!_{\Lambda}$Y
hypernuclei produced in the $(\pi^+,K^+)$ reaction on $^{51}$V and
$^{89}$Y targets at the beam momentum of $1.05\,$GeV. In drawing the smooth
spectral distributions, each level is a convolution with the appropriate Gaussian
width $\Gamma$ depending on its character in the experimental data. We used
only a single Gaussian for all the levels in contrast to Ref.~\cite{hot01b},
where more than one Gaussian were used for some levels.

In $^{51}\!\!\!_{\Lambda}$V case the series of levels are obtained by
the following configurations and binding energies ($E_{\text{bind}}$):
$[(f_{7/2}^{-1})_N, (s_{1/2})_\Lambda]$ ($3^-$,
$E_{\text{bind}} = 19.75\,$MeV), $[(f_{7/2}^{-1})_N, p_\Lambda]$ ($4^+$, $E_{\text{bind}} = 11.75\,$MeV),
and $[(f_{7/2}^{-1})_N, d_\Lambda]$ ($5^-$, $E_{\text{bind}} = 3.75\,$MeV). These states
make the largest contributions to the corresponding cross sections. On the
other hand, the levels in $^{89}\!\!\!_{\Lambda}$Y are obtained with
the configuration where the neutron hole state corresponds to the
$g_{9/2}$ orbit with the $\Lambda$ in the $0s$, $0p$, $0d$ and $0f$ orbitals.
The configurations of these levels are:
$[(g_{9/2}^{-1})_N, (0s)_\Lambda](4^+, E_{\text{bind}} = 23.6\,\text{MeV})$,
$[(g_{9/2}^{-1})_N, (0p)_\Lambda](5^-, E_{\text{bind}} = 16.5\,\text{MeV})$,
$[(g_{9/2}^{-1})_N, (0d)_\Lambda](6^+, E_{\text{bind}} = 10.0\,\text{MeV})$,
$[(g_{9/2}^{-1})_N, (0f)_\Lambda](7^-, E_{\text{bind}} = 2.3\,\text{MeV})$.

We see that for both nuclei the calculated spectral distributions
reproduce the overall global trends of the data of
Ref.~\cite{hot01b} reasonably well. Our calculations reproduce the experimentally observed
three and four clear peaks in the $^{51}\!\!\!_{\Lambda}$V
and $^{89}\!\!\!_{\Lambda}$Y spectra, respectively. However, looking more closely
we note that the theory overestimates the experimental
cross sections for the $(0s)_\Lambda$ orbitals somewhat in both
cases. For these states, the corresponding binding energies
($E_{\text{bind}}$) are quite large and for these strongly mismatched cases,
distortion effects could play a more significant role.

In both cases, one observed some minor peaks which fill up the gaps
between the major peaks.
These peaks correspond to core excited states and the
states corresponding to other, less dominant, members of the
configurations mentioned above. We have not made any attempt to fit to
these peaks in this exploratory study. Clearly, a quantitative
description of the spectral shapes of the heavier mass $\Lambda$
hypernuclei requires proper consideration of the core excitation
and mixing of states of different parity.

\subsection{Effects of initial and final state interactions }
\label{chap:res.dw}

The calculations presented thus far have been done in a plane wave
approximation, where the pion-nucleus and kaon-nucleus interactions
are ignored. While the essential features of the high momentum
transfer reaction $(\pi^+,K^+)$ can be understood in this approach,
the nuclear interactions may have some consequences. They produce
both absorptive and dispersive effects Ref.~\cite{dov80}. However, for
the large incident energies considered in this calculation, the
absorption effects are likely to be the more important.

One should also note that in the distorted wave
treatment the continuum wave functions are no longer associated with
sharp momenta but are states with
a momentum distribution, see Eqs.~\eqref{eq:mesonwf.pi}, and~\eqref{eq:mesonwf.ka}.
This leads to a redistribution of the momentum transfer
differently from what is allowed in the plane wave approximation.
It could shift the sensitivity of the model to even lower momenta
leading to enhanced cross sections.
The competition between this effect and the absorption effect would
ultimately decide the role of distortions in these reactions.

We have estimated the role of the distortion effects in the initial
channel on the absolute magnitude of the cross sections. For this purpose,
we make use of the eikonal approximation as discussed in section~\ref{sec:mes.pie}.
Cross sections have been calculated in both PW and eikonal approximation
at one most forward angle for $^{12}$C$(\pi^+,K^+)^{12}\!\!\!\!\!_\Lambda$C,
$^{40}$Ca$(\pi^+,K^+)^{40}\!\!\!\!\!_\Lambda$Ca and
$^{89}$Y$(\pi^+,K^+)^{89}\!\!\!\!\!_\Lambda$Y reactions.
The results are shown in table~\ref{tab:pw.dw.comp}.
\begin{table}
	\caption{The differential cross sections at selected angles for
	the plane wave approximation and the distorted wave calculations for
	the pion in the eikonal approximation.}\label{tab:pw.dw.comp}
	\centering
	\begin{tabular}{ c c c c }
		\toprule
		 & &
		 \multicolumn{2}{c}{$\ud\sigma/\ud\Omega$\;[\textmu b/sr] } \\
		 nucleus & transition & plane wave & eikonal \\
		\midrule
		 $^{12}$C ($\theta_K = 1^\circ$) &
		 $n p_{3/2} \rightarrow \Lambda s_{1/2}$ &
		 16.55 & 13.21 \\
		 & $n p_{3/2} \rightarrow \Lambda p_{3/2}$ &
		 10.16 & 7.10 \\
		 \midrule
		 $^{40}$Ca ($\theta_K = 4^\circ$) &
		 $n d_{3/2} \rightarrow \Lambda s_{1/2}$ &
		 15.55 & 3.18 \\
		 \midrule
		 $^{89}$Y ($\theta_K = 4^\circ$) &
		 $n g_{9/2} \rightarrow \Lambda s_{1/2}$ &
		 7.43 & 0.642 \\
		\bottomrule
	\end{tabular}
\end{table}

We see that the distortion effects in the incident channel lead to the
reduction of the peak cross sections by factors of $1.2$ to $12$ as the target
mass varies from $12$ to $89$. Our result for the $^{12}$C target is
in contrast to that of the non-relativistic impulse model of Ref.~\cite{dov80},
where the distortion effects were found to reduce the differential cross
section by about an order of magnitude for this target. The distortion
effects play an increasingly important role with increasing mass of the
target nucleus. This result is in agreement with those of Refs~\cite{shy06},
and ~\cite{Bennhold:1989bw}.

\section{Summary and conclusions}
\label{chap:summary}

In summary, we studied the $A(\pi^+,K^+)_{\Lambda}A$ reaction on
$^{12}$C, $^{40}$Ca, $^{51}$V and $^{89}$Y targets within a fully
covariant model, where in the initial collision of incident pion
with one of the target nucleons the $N^*$(1650), $N^*$(1710), and
$N^*$(1720) intermediate baryon resonance states are excited
which subsequently propagate and decay into the relevant channel.
We have retained the full field theoretic forms of various
interaction vertices and obtained the baryon bound states
by solving the Dirac equation with appropriate scalar and vector
potentials. The vertex constants were taken to be the same as those
determined in previous studies. We have ignored, for the time being,
the distortion effects in the incident and outgoing channels as our
main aim in this paper has been to establish a fully covariant model
for this reaction which has so far been described only within a
non-relativistic distorted wave impulse approximation picture. The
plane wave approximation facilitates the application of this
novel approach to very many cases without requiring lengthy and
cumbersome computations which are necessarily involved in the
distorted wave methods.

We find that excitations of $N^*$(1650), and $N^*$(1710) resonant
states dominate the cross sections for the $(\pi^+,K^+)$ reaction
for beam energies below $2\,$GeV. Our model describes well the shapes
of the experimental angular distributions for two states which
corresponds to the prominent peaks in the $^{12}\!\!\!_{\Lambda}$C
spectrum. The total cross sections show a substantial dependence on
the beam energy with a distinct peak around $1\,$GeV for reactions on
both $^{12}$C and $^{40}$Ca targets. The differential cross sections
peak near zero degrees for both light mass as well as heavier
targets. Thus, measurements at forward angles and at beam energies
around $1.0\,$GeV are expected to have high yields.

The characteristic bump structures reflecting the $\Lambda$
major shell orbits as seen in the spectra of the
$^{51}\!\!\!_{\Lambda}$V and $^{89}\!\!\!_{\Lambda}$Y hypernuclei
were reproduced reasonably well by our model. However, for a
quantitative description of the spectra, more realistic
calculations with configuration mixed $\Lambda$ particle, neutron
hole states are required. Mixing of the different parity
states within a given particle-hole configuration may also be
needed.

Our work shows that it is indeed feasible to have a  fully covariant
description of the hypernuclear production via the $(\pi^+,K^+)$ reaction.
In future studies, the structure aspects of the model should be
improved. Also, alternative covariant bound state wave
functions~\cite{kei00b,shy09b} should be used in order to check the
validity of the method used in this work to determine them.
Distortion effects in the initial and final channels are required
in order to have more reliable predictions of the absolute
magnitudes of the cross sections.

\section{ACKNOWLEDGEMENTS}
This work has been supported by GSI Darmstadt, the
Sonderforschungsbereich/Transregio 16, Bonn--Giessen--Bochum of the
German Research Foundation (DFG), and the HIC for FAIR LOEWE centre.

\end{document}